\documentclass[10pt,twocolumn]{article}
\usepackage[top=1.9cm, bottom=1.9cm, left=1.8cm, right=1.8cm]{geometry}
\usepackage{times}  %
\usepackage[sort&compress,numbers]{natbib}  %
\usepackage[hyphens]{url}
\usepackage{graphicx}  %
\usepackage[hyphens]{url}  %
\usepackage{graphicx} %
\usepackage{tikzsymbols}

\newcommand{\descr}[1]{\smallskip\noindent\textbf{#1}}

\usepackage{sectsty}
\sectionfont{\bfseries\Large\raggedright}

\usepackage{url}                                %
\usepackage{array,multirow,graphicx,adjustbox}  %
\usepackage{booktabs}                           %
\usepackage[utf8]{inputenc}
\usepackage[ruled,algosection,noend,linesnumbered]{algorithm2e}
\usepackage{float}
\usepackage{paralist}
\usepackage{amsmath}
\usepackage{amsfonts}
\usepackage{xcolor}
\usepackage{csquotes} 
\usepackage{tabularx}
\newcolumntype{L}[1]{>{\raggedright\arraybackslash}p{#1}}

\usepackage{makecell}
\usepackage{pbox}
\usepackage[hang,flushmargin,symbol]{footmisc}
\usepackage{flushend}
\usepackage{xspace}
\usepackage{multicol, lipsum}
\usepackage[T1]{fontenc}

\usepackage{xcolor}
\usepackage{booktabs}
\usepackage{graphicx}
\usepackage{paralist}
\usepackage[bf]{caption}
\usepackage{subcaption}

\let\oldbibliography\thebibliography
\renewcommand{\thebibliography}[1]{%
  \oldbibliography{#1}%
  \setlength{\itemsep}{2pt}%
}

\usepackage[hang,flushmargin]{footmisc}

\usepackage[compact]{titlesec}
\titlespacing*{\section}{0pt}{*4}{4pt}
\titlespacing*{\subsection}{0pt}{*2.5}{2.5pt}
\usepackage{xspace}

\makeatletter
\def\url@leostyle{%
  \@ifundefined{selectfont}{\def\UrlFont{}}%
  {\def\UrlFont{}}%
}
\makeatother
\usepackage[hyphenbreaks]{breakurl}

\usepackage[bookmarks=true, bookmarksnumbered=true, colorlinks=true, linkcolor=linkcol, citecolor=citecol, urlcolor=urlcol, hypertexnames=true]{hyperref}

\definecolor{darkgreen}{RGB}{0, 100, 0}
\definecolor{linkcol}{rgb}{0.3,0,0}
\definecolor{citecol}{rgb}{0.3,0,0}
\definecolor{urlcol}{rgb}{0.3,0,0}

\makeatletter
\def\url@leostyle{%
  \@ifundefined{selectfont}{\def\UrlFont{\small}}%
  {\def\UrlFont{}}%
}
\makeatother
\usepackage[capitalise]{cleveref}

\newcommand{\jbnote}[1]{}
\newcommand{\synote}[1]{}
\newcommand{\edc}[1]{}

\begin{document}

\sloppy

\title{\bf Beyond Fish and Bicycles: Exploring the Varieties of Online Women's Ideological Spaces\footnotemark}

\author{Utkucan Balcı$^1$, Chen Ling$^2$, Emiliano De Cristofaro$^3$,\\Megan Squire$^4$, Gianluca Stringhini$^2$, Jeremy Blackburn$^1$\\[1ex]
$^1$Binghamton University, $^2$Boston University, $^3$University College London, $^4$SPLC}

\date{}

\maketitle

\begin{abstract}
  
  The Internet has been instrumental in connecting under-represented and vulnerable groups of people.
  Platforms built to foster social interaction and engagement have enabled historically disenfranchised groups to have a voice.
  One such vulnerable group is women.
  In this paper, we explore the diversity in online women's ideological spaces using a multi-dimensional approach.
  We perform a large-scale, data-driven analysis of over 6M Reddit comments and submissions from 14 subreddits.
  We elicit a diverse taxonomy of online women's ideological spaces, ranging from counterparts to the so-called Manosphere to Gender-Critical Feminism.
  We then perform content analysis, finding meaningful differences across topics and communities.
  Finally, we shed light on two platforms, ovarit.com and thepinkpill.co, where two toxic communities of online women's ideological spaces (Gender-Critical Feminism and Femcels) migrated after their ban on Reddit.

\end{abstract}

\renewcommand{\thefootnote}{\fnsymbol{footnote}}
\footnotetext[1]{\noindent Published in the Proceedings of the 15th ACM Web Science Conference 2023 (ACM WebSci 2023). Please cite the WebSci version.}

\renewcommand*{\thefootnote}{\arabic{footnote}}

\section{Introduction}
\label{sec:intro}

While issues of misogyny and the LGBTQ+ community have been a part of public discourse for quite some time, the focus on \emph{gender}, and specifically women, has become central. %
Although gender studies is a well-established field, discussions of gender identity have grown well beyond academic circles.
In particular, online communities focused on gender identity of women have emerged, birthing full-blown social movements.

In this paper, we perform the first large-scale, data-driven study of online women's ideological communities -- i.e., communities that are focused on not just issues that women face but also what it means to be a woman.
To do so, we construct a dataset of over 6 million posts/comments from 14 women-oriented ideological subreddits.
We perform a data-driven analysis across several axes, focusing on identifying and understanding different subcommunities within our dataset.

First, we identify women-oriented ideological subreddits and quantitatively derive a taxonomy from them, resulting in three clusters: 1)~Mainstream Feminism, 2)~Gender-Critical Feminism, and 3)~Manosphere Analogs.\footnote{The Manosphere is, roughly speaking, a conglomerate of Web-based, mostly misogynist, movements focused on ``men’s issues''~\cite{lilly2016world,ribeiro2020evolution}.}
We compare the user-base similarity of each community, finding similiarties between Gender-Critical Feminism (which became very popular a few months before being banned) and two Manosphere Analog communities, Femcels and WGTOW. %

We then perform a quantitative topic analysis using Top2Vec, a model extracting semantic meaning from text. %
We also analyze, qualitatively, a random sample of 100 posts from the various topics; this reveals how even discussions about mundane topics like clothing can serve to differentiate the communities.
We find that Mainstream Feminism is more focused on women's problems related to toxic masculinity; also, among Manosphere Analogs, the overall leaning of the posts of Female Dating Strategy, Femcels, and WGTOW are more similar to each other compared to Red Pill Women. 

Finally, we explore the effect of the deplatforming of two toxic communities in our dataset, Gender-Critical Feminism and Femcels.
We find that bans increased the toxicity of latter and decreased that of the former. %
That said, deplatforming appears to have \emph{increased} identity attacks in both communities.

\section{Background \& Related Work}
\label{sec:background}

Gender-based ideologies and movements are much older than the Internet and social media.
First-wave feminism dates back to the 19th century~\cite{henry2004not}, the transgender rights movement to the 1950s~\cite{barton2020oral}, and the men's rights movement to 1970s~\cite{messner1998limits}.
Although there are studies to understand gendered online movements and communities~\cite{khan2020reddit,lucy2018using}, these mostly lack a variety of online women's ideological spaces, prompting the need for a broader taxonomy. 

\descr{Online Feminism.}
Feminism, in general, is defined as ``the belief in social, economic, and political equality of the sexes~\cite{brunell}.''
In the past, studies on online feminism have predominantly examined the movement as a cohesive whole, without differentiating between its various branches or perspectives.
Previous studies have examined the use of hashtags on Twitter to gain insights into online feminist movements~\cite{loza2014hashtag,dixon2014feminist,mendes2018metoo,zimmerman2017intersectionality, worthington2020celebrity, antunovic2019we}.
Khan and Golab~\cite{khan2020reddit} analyzed gendered movements on Reddit, focusing on r/Feminism, r/MensRights, and r/TheRedPill; they found that users of r/Feminism were more similar to those of the Men's Liberation Movement, a feminist ally, than users of r/MensRights and r/TheRedPill.
Similarly, Brattland~\cite{brattland2017information} examined the information preferences of users on r/Feminism and r/MensRights using r/News to investigate the Brock Turner case, and found that only the former actively worked to combat rape culture.
Simoes et al.~\cite{simoes2021new} argue that online feminism and misogyny gain strength under similar online conditions.

Studies examining feminist Facebook pages have found that they offer greater participation and empowerment to feminist activism, serving as a safe space~\cite{mclean2016learning,flores2020reviving,clark2018building}, and feminism is also linked to greater internal political efficacy~\cite{heger2021feminism}.
Moreover, while feminism uses ``trolling'' to celebrate women's embodiment and sexuality and to engage with feminist values, feminists do not suppress debate when being trolled~\cite{massanari2019come, herring2002searching}.

These studies do not always distinguish between its various branches, even though prior research has identified a division within feminism based on either liberal versus radical or trans-inclusive versus trans-exclusive viewpoints~\cite{serrano2021nlp,willem2021gender,hines2019feminist}.
Gender-Critical Feminism (GCF), also derogatorily known as trans-exclusionary radical feminism (TERF), views transgender people's activism and visibility as a possible threat to women's rights and spaces due to their challenging of the binary understanding of sex and gender~\cite{jones2020toilet,simon2021isn}.
In June 2020, Reddit banned r/GenderCritical in response to new hate speech rules, leading to the migration of r/GenderCritical to a Reddit-like platform called ovarit.com~\cite{tiffany2020}.
Previous research used Twitter data to build classifiers to detect TERF hate groups on Twitter~\cite{lu2020computational}.
Nonetheless, there has not been any large-scale, comprehensive study that analyzes and compares the more liberal, trans-inclusive side of feminism and the more radical, trans-exclusive side of feminism.

\descr{Manosphere ``Analogs.''}
One well-explored gender-based ideological space is a collection of men's ideological communities called ``the Manosphere.''
Lilly~\cite{lilly2016world} presented a taxonomy of the Manosphere by analyzing its subcultures, showing these communities view masculinity to be ``in crisis,'' constantly under attack by feminizing forces, and see feminism as hypocritical and repressive.   
Communities within the Manosphere exhibit disturbingly increased levels of toxicity, hate, and a proclivity for real-world violence~\cite{ribeiro2020evolution,farrell2019exploring,copland2020reddit}.
There is also a connection between them and the alt-right~\cite{mamie2021anti,adams2021augmentation,vu2021extremebb}.  
The alt-right has embraced the practice of describing women as ``shield maidens'' and ``trad wives,'' which serves to mask their underlying white supremacist ideology~\cite{love2020shield}.

The Manosphere, due to its nature, has a notably low participation rate among women~\cite{squire2019way}.
Nevertheless, its increasing popularity has led to the emergence of women's communities analogous to those in the Manosphere.
One example of these ``Manosphere Analogs'' is femcels (female involuntary celibates).
As the analog of incels, the women in this community cannot find a sexual or a romantic partner, and many share their experiences of living with a body that is deemed unappealing to men~\cite{fong2022,van2020analysing}.
Similar to other \emph{pilled} ideologies, femcels refer to their ``realization'' about society as taking the \emph{pink pill}~\cite{tiffany2022,colombo2022}.
In January 2021, a femcel subreddit, r/Trufemcels, was banned on Reddit for violating the site's rules against promoting hate, leading to the community's migration to thepinkpill.co as an alternative~\cite{aronowitz2022}.
Another example is Red Pill Women (RPW).
As the analog of the Manosphere subculture The Red Pill~\cite{dignam2019misogynistic}, most of the users in this community do not believe in gender equality, and they discuss house-wife duties, or supporting their alpha men~\cite{mcgurran2014bitter,dignam2019misogynistic}. 
Even though there are previous studies on RPW and femcels~\cite{ling2022femcels,lucy2022discovering,van2020analysing,zdjelar2020alone, labbaf2019united, mcgurran2014bitter,jarvis2021disentangling,brix2023lack}, there have not been any large-scale studies focusing on them.

There are also understudied Manosphere Analog communities, or communities that have been covered in online journalistic articles, but only recently gained attention from the academic research community.
One such community is Female Dating Strategy (FDS), present on Reddit as r/FemaleDatingStrategy, and seemingly similar to a Manosphere subgroup called Pick Up Artists (PUA). 
Like PUA, FDS members objectify the opposite gender by approaching interactions as some kind of a game~\cite{taylor2020,scott2020,sisley2021}.
Ling~\cite{ling2022femcels} examined a sample of r/FemaleDatingStrategy, considering it as a movement related to femcels, but we treat it as a distinct community due to their similarities to PUA.
Notably, Ling identified parallels between femcel ideology and radical feminism.
Another study~\cite{lucy2022discovering} found that Femcels and FDS use certain words in an analogous manner to Manosphere communities.
Another community is WGTOW (Women Going Their Own Way).
As the analog of Men Going Their Own Way (MGTOW), this community consists of \emph{women} who aim to separate their lives with men, and they have gathered on r/wgtow on Reddit~\cite{miles2022}.

\section{Dataset}
\label{sec:dataset}

While previous research investigated various aspects of online women's ideological spaces (OWIS)~\cite{khan2020reddit,kay2021abject,mcgurran2014bitter,bergstrom2015feminism}, a well-defined taxonomy is currently lacking, and certain communities within OWIS remain unexplored.
To address this gap, we select a variety of feminist and Manosphere Analog subreddits as seeds; then, we identify subreddits with a similar community structure (i.e., subreddits that are a part of OWIS).

\subsection{Identifying Women-Based Ideological Subreddits}
We set out to build a compact cluster that includes all of our seed subreddits, which consist of women-based ideological subreddits covering many viewpoints (e.g., feminist and anti-feminist). 

\descr{Feminist Seed Subreddits.}
We select r/Feminism, which is previously used in studies to represent feminism~\cite{khan2020reddit,jacobs2020reddit,malik2022capturing, brattland2017information}, as well as r/GenderCritical.
While GCF has a relatively toxic culture and has made alliances with non-feminist groups that share their anti-trans views~\cite{burns2019}, it is a feminist community nevertheless.
By including both r/Feminism and r/GenderCritical as seeds, we aim to capture a wider portion of the OWIS spectrum.

\descr{Manosphere Analog Seed Subreddits.}
To represent Female Dating Strategy, Femcels, Red Pill Women, and WGTOW, we choose r/FemaleDatingStrategy, r/Trufemcels, r/RedPillWomen, and r/wgtow, as their relationships to these communities are explained in Section~\ref{sec:background}.

\descr{Subreddit Embeddings.}
To identify the subreddits that belong to OWIS, we rely on pre-trained subreddit embeddings provided by Raymond et al.~\cite{raymond2022measuring}, who build on~\cite{waller2021quantifying} to train subreddit embeddings (aiming to explore the community structure of Reddit).
This embedding is trained with word2vecf~\cite{levy2014dependency}, using users as \emph{contexts} and subreddits as \emph{words}.
The training data consists of comments posted to the top 10,000 subreddits of 2019, which encompass 95.6\% of all comments made that year (1.59 billion comments from 19.34 million users)~\cite{raymond2022measuring}. 

We opt for this embedding for three reasons:
1)~the seed subreddits we select were active between February 2019 (the first post of r/FemaleDatingStrategy) and June 2020 (the last post of r/GenderCritical before its ban), and 2019 accounts for almost two-thirds of this time period,
2)~of the six seed subreddits, five are among the 10,000 subreddits featured in this embedding, with the exception of r/wgtow,
and 3)~as we show in Section~\ref{sec:user_analysis}, the normalized user overlaps of each subreddit we identify later in this section exhibit a strong correlation with their corresponding normalized user overlaps from 2019.

\descr{Clustering.}
After applying dimensionality reduction with UMAP~\cite{mcinnes2018umap}, we perform agglomerative clustering~\cite{murtagh2012algorithms}, which is hierarchical clustering that follows a bottom-up approach by merging the closest clusters at each step, with the number of clusters controlled by a parameter ($k$).
We start with $k = 10$ and increment it by 10 until we find the largest $k$ that results in all of the seed subreddits appearing in the same cluster (i.e., while there is still a single, potential OWIS cluster).
This method produces 50 clusters, where the cluster containing our seed subreddits has a size of 284.
Upon manual inspection, we find these subreddits cover a broader set of topics than just OWIS, including dating-related subreddits (e.g., r/Tinder, r/Bumble, r/dating\_advice), self-improvement subreddits (r/selfimprovement, r/IWantToLearn, r/getdisciplined), men's ideological online spaces (e.g., r/TheredPill, r/GenderCriticalGuys, r/MensRights), and many general interest subreddits (r/AmItheAsshole, r/confession, r/offmychest).
To get a more compact cluster, we apply another step of clustering.
We use the same technique on the vectors of the subreddits of this cluster, starting from $k = 1$ and incrementing by one until we get the largest cluster that contains all seed subreddits.
With $k = 7$, we get 69 subreddits, most of which are gender-based ideological spaces or subreddits based on appearance (e.g., r/short, r/tall, r/amiugly).
We believe these subreddits are clustered with gender-based ideological spaces mainly because of their relations with Incel subreddits (e.g., r/shortcels).

\descr{Results.}
To extract women-based ideological subreddits, we manually examine the subreddits in this cluster in terms of names, titles, bios, and banners of these subreddits, choosing the ones that are clearly women-oriented according to our interpretation of the relevant literature (i.e., self-statements that they are women oriented or belong to a women-oriented community).
Overall, we identify 15 women-based subreddits, but ultimately exclude two that are based on appearance (r/TallGirls and r/GirlsMirin), as we find no evidence of them being ideology related.
In the end, our dataset includes 14 subreddits, with 13 from the clustering, along with r/wgtow, a clear analog of MGTOW.

\begin{table}[t]
  \centering
  \small
  \setlength{\tabcolsep}{4pt}

  \begin{tabular}{lrrr}
      \toprule
      {\bf Subreddit} & {\bf \# Posts} & {\bf \#Authors} & {\bf Min--Max Date}   \\ 
      \midrule
      r/FemaleDatingStrategy        & 1,763,702                    & 48,243                         & 02/19 - 01/22                                  \\
      r/GenderCritical              & 1,629,169                    & 55,649                         & 09/13 - 06/20                                  \\
      r/AskFeminists                & 937,716                      & 59,155                         & 07/11 - 01/22                                \\
      r/Feminism                    & 779,887                      & 140,615                        & 01/09 - 01/22                                 \\
      r/RedPillWomen                & 395,097                      & 27,669                         & 06/13 - 01/22                                  \\
      r/Trufemcels                  & 240,934                      & 12,936                         & 04/18 - 01/21                                  \\
      r/women                       & 267,011                      & 52,331                         & 02/08 - 01/22                                  \\
      r/AskTruFemcels               & 215,418                      & 5,884                          & 10/18 - 01/21                                 \\
      r/Vindicta                    & 183,394                      & 16,037                         & 08/19 - 01/22                                  \\
      r/itsafetish                  & 155,359                      & 13,385                         & 08/18 - 06/20                                  \\
      r/PinkpillFeminism            & 138,098                      & 7,609                          & 06/19 - 01/21                                 \\
      r/TrollGC                     & 69,847                       & 5,880                          & 03/17 - 07/20                                  \\                                                     
      r/TheGlowUp                   & 65,350                       & 11,310                         & 03/19 - 01/22                                 \\
      r/wgtow                       & 35,644                       & 4,026                          & 05/14 - 01/22                                  \\
      \midrule
      \bf All                           & \bf 6,876,566                    & \bf 394,315               & \bf 02/08 - 01/22                                  \\
      & & (Unique)\\
      \bottomrule    
  \end{tabular}
  \caption{Overview of the subreddits in our dataset.}
\label{tab:clusters_info}
  \end{table}

\subsection{Data Collection}
We collect 6,876,566 posts (submissions + comments) from the 14 subreddits using the Pushshift API~\cite{baumgartner2020pushshift} between February 2008 and February 2022.
Table~\ref{tab:clusters_info} reports the total number of posts, and the total number of authors, with the earliest and latest post dates of each subreddit.
The most popular subreddit is r/FemaleDatingStrategy with 1,763,702 posts, followed by r/GenderCritical with 1,629,169 posts, while r/wgtow is the smallest, with 35,644 posts.

\descr{Ethics.}
We are aware that social media data can potentially reveal personal information. 
To ensure that our research adheres to ethical principles, we adopt standard best practices~\cite{bailey2012menlo,rivers2014ethical}, including not attempting to further de-anonymize any author.
As our study solely employs publicly available data and does not involve interactions with participants, our institution does not consider it as human subjects research.
However, to avoid targeting any user on the basis of their comments featured in our topic analysis, when presenting examples, we paraphrase text while preserving its original tone and context.

\begin{figure*}[t]
  \centering
  \includegraphics[width=\textwidth]{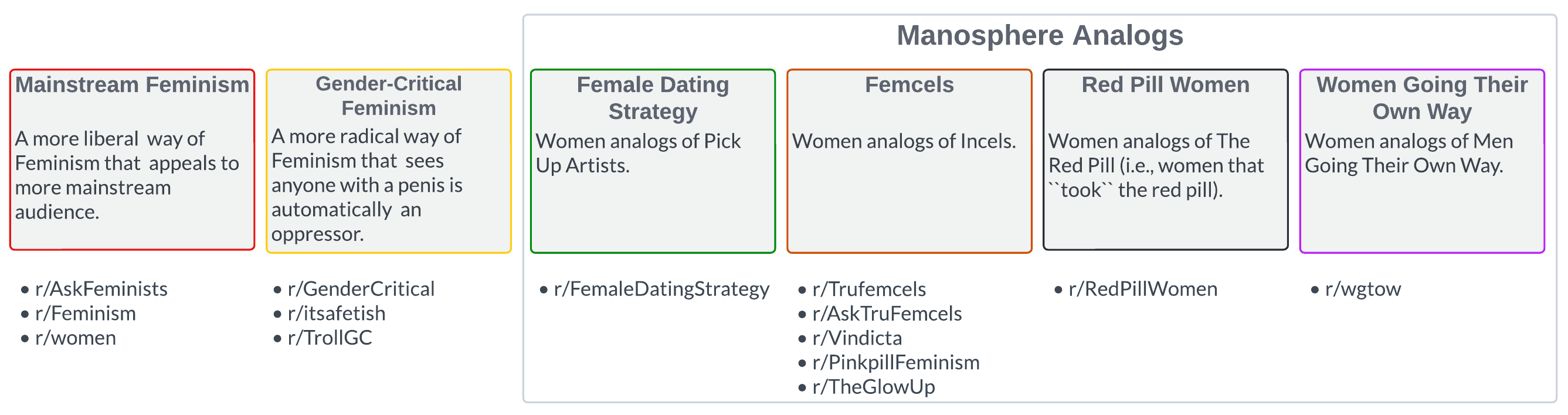}
  \caption{Diagram of the taxonomy of online women's ideological spaces.}
  \label{fig:diagram}
\end{figure*}

\section{User Analysis}
\label{sec:user_analysis}

In this section, we present our taxonomy of OWIS. 
Then, we analyze the activity and the user-bases of the communities we identify within the taxonomy.

\subsection{Creating an OWIS Taxonomy}

In Section~\ref{sec:dataset}, we use agglomerative clustering to find a cluster that keeps all our seed subreddits by controlling the cluster size.
Next, to create a taxonomy, we perform agglomerative clustering without controlling the cluster size to automatically find the communities of OWIS.  %
In the end, we find three main clusters: 
1)~Mainstream Feminism.
2)~Gender-Critical Feminism (GCF),
and 3)~Manosophere Analogs.
For the latter, rather than examining all the subreddits together, we split the cluster into their respective Manosphere Analog communities (Femcels, FDS, and RPW) while treating r/wgtow as its own community (due to being a direct analog of WGTOW) to allow for more fine-grained analysis.

Figure~\ref{fig:diagram} presents our final taxonomy, which goes to complement Table~\ref{tab:clusters_info}.
Note that Mainstream Feminist subreddits are the oldest, and have the most posts.
Also, due to being banned from Reddit, three of the five Femcel subreddits and all of the GCF subreddits most recent posts are in 2020 and 2021, about a year before the end of our dataset.

\begin{figure}[t]
	\centering
	\includegraphics[width=0.95\columnwidth]{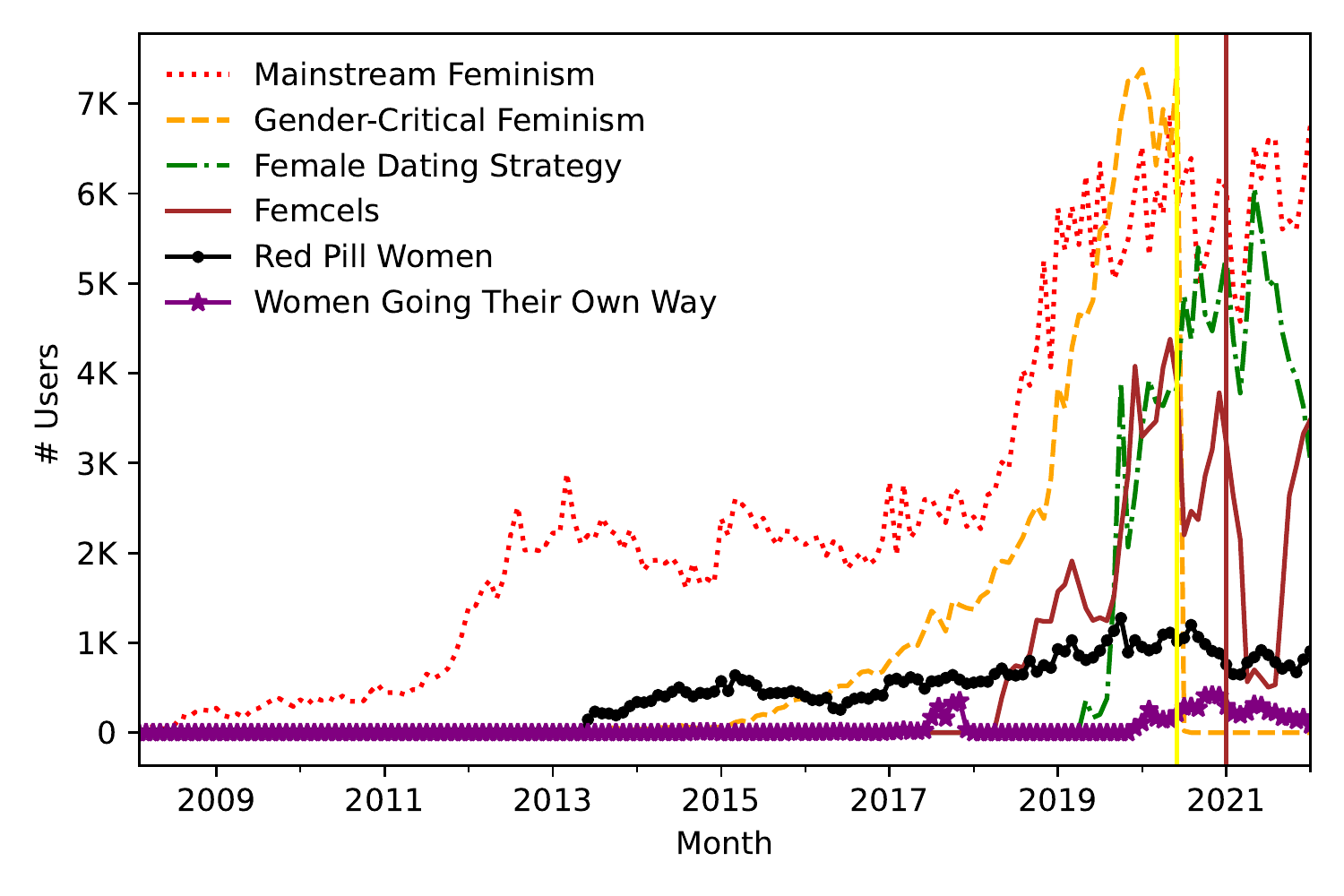}
	\caption{Monthly user counts of each community in our dataset. 
	The yellow vertical line corresponds to r/GenderCritical being banned, and the brown one when r/Trufemcels, r/AskTruFemcels, and r/PinkpillFeminism were.
}
	\label{fig:monthly_user}
\end{figure}

\subsection{Monthly User Activity}
To understand the popularity of the communities over time, we look for their monthly user counts.
Figure~\ref{fig:monthly_user} shows Mainstream Feminism is the most popular community for the majority of the time.
We find GCF has become the most popular community in 9 out of 10 months in their last year before their ban on Reddit. 
We also see a sharp increase in FDS after its establishment, where they became the most popular community of September 2019.
We observe a trend of decrease in FDS starting after January 2021.
Although the decrease in the monthly user counts of Femcels is naturally expected after three of their subreddits were banned on January 2021, we also see a decrease in this community after the ban on GCF subreddits.
As we explore it next, this can indicate an underlying relationship between the authors of GCF and Femcels.
For RPW, we see a relatively stable monthly user count, with an average of 616 monthly users.
As the smallest community, WGTOW has the fewest monthly users, averaging 114.

\begin{figure*}[t]
	\centering
	\includegraphics[width=1.25\columnwidth]{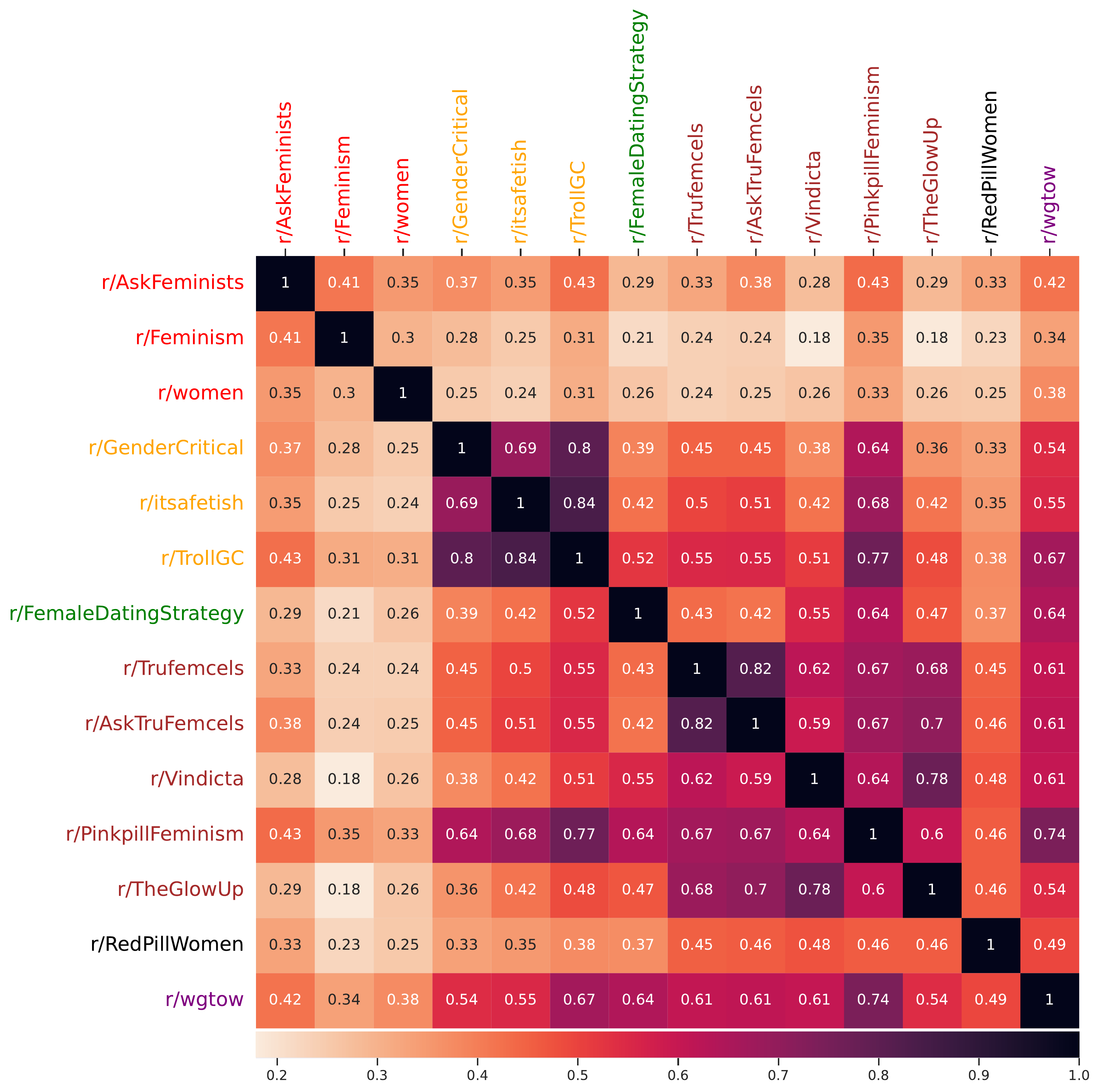}
	\caption{Normalized user overlaps between each community. Darker color/higher score means that the user-bases are more similar. GCF is more similar to the six Manosphere Analog subreddits than Mainstream Feminist subreddits. }
	\label{fig:normalized_user_overlaps}
\end{figure*}

\subsection{User-Base Overlaps}
To look deeper into the relationship between each community, we examine the user overlaps between each subreddit.
Since there are different amounts of users in each community, following Squire's work~\cite{Squire2019NetworkTA}, we normalize the user overlaps with Bonacich's method~\cite{bonacich1972technique}.
From Figure~\ref{fig:normalized_user_overlaps}, we see Mainstream Feminism has the least inter-community normalized user overlaps (mean: 0.35) compared to the other two communities with more than one subreddit, namely, GCF (mean: 0.77) and Femcels (mean:0.67).
This might be due to the subreddits of Mainstream Feminism appealing to more \emph{mainstream} audiences than the other OWIS communities.

We find r/GenderCritical is less similar to r/Feminism compared to the two subreddits of Manosphere Analogs, r/PinkPillFeminism and r/wgtow.
Considering r/GenderCritical has a more normalized user overlap score with 6 out of 8 Manosphere Analog subreddits compared to any Mainstream Feminist subreddit, this might imply that GCF is a fringe feminist community.
Moreover, RPW is more similar to Femcels and WGTOW compared to other communities.
Also, despite WGTOW having a 0.74 normalized user overlap with r/PinkpillFeminism (a Femcel subreddit), they are closer to FDS compared to their normalized mean overlaps with Femcel subreddits (0.62).

Since we use subreddit embeddings from 2019 to create our taxonomy, we also check whether the user-base overlaps from 2019 correlate with our results.
To this end, we calculate Spearman correlation, finding it to be strong ($\rho = 0.95$, $p < 0.01$).

\descr{Takeaways.}
Mainstream Feminism, GCF, and Manosophere Analogs are the three main communities of OWIS.
Interestingly, even though Mainstream Feminism was the most popular community of OWIS for most of the time, GCF overtook it closer to their ban on Reddit.
We also show that r/GenderCritical has more normalized user overlaps with 6 out of 8 Manosphere Analog subreddits compared to any of the Mainstream Feminist subreddits; this suggests an underlying relationship with GCF and some of the Manosphere Analogs.

\section{Topic Analysis}
\label{sec:topic_analysis}

We now we examine the main focus of each community by analyzing their most popular topics, which allows us to study, at scale, differences in discussion across the communities.

\descr{Topic Modeling.} We train a Top2Vec model to explore the most popular topics of the communities of OWIS.
Top2Vec finds the topics of the documents by clustering their embedding vectors using UMAP and HDBSCAN~\cite{mcinnes2017hdbscan}, where this intuition comes from the assumption of many semantically similar \emph{documents} (in our case, posts on Reddit) are indicative of an underlying topic~\cite{angelov2020top2vec}.
We choose Top2Vec over ``traditional'' topic modeling techniques like LDA for three main reasons: 
1)~Top2Vec does not ignore the ordering and semantics of words, and makes it possible to automatically find the number of topics~\cite{angelov2020top2vec};
2)~It finds topics based on documents instead of words; and
3)~Its underlying techniques help us capture more specific topics, and yield a better intuition about each community. 

While training our model, we use default parameters, remove hyperlinks, and the posts from r/AutoModerator to avoid auto-moderation posts, and train a Doc2Vec~\cite{le2014distributed} model from scratch to ensure unique vocabulary (e.g., slang) is captured.
We then conduct a more in-depth analysis of their popular topics by manually reviewing 100 sampled posts for the top 10 most frequently discussed topics in each community.
We remove at most one topic for each community if we consider it an outlier consisting of incoherent posts with basic replies to other posts or salutations.
\cref*{tab:feminism_top10,tab:gcf_top10,tab:fds_top10,tab:femcels_top10,tab:rpw_top10,tab:wgtow_top10} show the top 10 topics of each community.

\begin{table}[t]
  \small
  \centering
    \setlength{\tabcolsep}{3pt}
    \begin{tabularx}{\columnwidth}{rL{5.3cm}L{1.9cm}}
      \toprule
{\bf Size} & {\bf Top 5 Words}                               & {\bf Label}  \\ 
      \midrule
10,418                       & accuser, accusers, convict, allegation, convicting              & Sex Conviction    \\
9,904                        & salaries, promotions, graduates, fields, secretarial            & Career              \\
9,050                        & kyriarchy, oppressions, advantaged, systemically, axis          & Kyriarchy                  \\
8,580                        & hillary, presidential, presidency, primaries, biden             & Elections             \\
8,558                        & terminate, abort, aborts, adoption, lps                         & Abortion                        \\
7,921                        & intoxicated, inebriated, inebriation, drunkenness, intoxication & Sexual Assault                \\
7,327                        & brigaded, brigading, mensrights, brigades, srsers               & MRA (Reddit)                     \\
6,500                        & masculinity, manliness, feminity, unmanly, toxically            & Toxic Masculinity \\
6,499                        & mrm, masculism, egalitarians, menslib, mensrights               & MRA  \\
6,222                        & fetus, foetus, embryo, unborn, fetuses                          & Abortion (Fetus)          \\
     
      \bottomrule    
  \end{tabularx}
  \caption{Top 10 Topics of Mainstream Feminism.}
\label{tab:feminism_top10}
  \end{table}

\begin{table}[t]
  \small
  \centering
    \setlength{\tabcolsep}{3pt}
    \begin{tabularx}{\columnwidth}{rL{5cm}L{2.3cm}}
      \toprule
{\bf Size} & {\bf Top 5 Words}                               & {\bf Label}  \\ 
      \midrule
        9,446                        & primaries, democrat, bernie, biden, sanders                       & Elections \\
        8,567                        & stalls, urinals, restrooms, stall, toilets                        & Bathroom Access \\
        7,922                        & competed, olympic, championships, athletes, championship          & Trans Athletes    \\
        6,608                        & blouse, trousers, blouses, sneakers, skirts                       & Clothing        \\
        6,323                        & leftist, leftists, socialists, centrist, centrists                & Leftism            \\
        5,672                        & pimps, johns, prostituted, trafficked, legalization               & Sex Work    \\
        5,459                        & bisexuals, biphobia, bisexuality, bisexual, biphobic              & Sexual Orientation               \\
        5,418                        & agp, agps, autogynephilia, fetishistic, autogynephiles            & Autogynephilia       \\
        4,706                        & xxy, chromosomes, chromosomal, xy, karyotype                      & Karyotype       \\
        3,904                        & dysphoria, dysphoric, dysmorphic, detransition, anorexics         & Gender Dysphoria       \\
        \bottomrule    
    \end{tabularx}
    \caption{Top 10 Topics of Gender-Critical Feminism.}
  \label{tab:gcf_top10}
    \end{table}

\begin{table}[t]
  \small
  \centering
    \setlength{\tabcolsep}{3pt}
    \begin{tabularx}{\columnwidth}{rL{5cm}L{2.45cm}}
      \toprule
{\bf Size} & {\bf Top 5 Words}                               & {\bf Label}  \\ 
      \midrule

      6,652                        & bumble, swiping, swipe, hinge, okc                      & Dating Apps     \\
      4,766                        & valentine, gifts, gift, valentines, christmas           & Valentines Day                \\
      4,231                        & moissanite, carat, ring, jeweler, cz                    & Jewelry                   \\
      4,171                        & foreplay, orgasm, orgasming, orgasms, orgasmed          & Orgasm                    \\
      4,095                        & dishes, dishwasher, laundry, cleaning, kitchen          & Housework              \\
      3,987                        & europeans, ethnically, indians, southeast, arab         & Ethnicity                              \\
      3,969                        & restaurant, coffee, restaurants, dessert, appetizers    & Restaurants   \\
      3,420                        & block, unblock, delete, unblocked, deleting             & Blocking/ Unblocking       \\
      3,408                        & cats, dogs, pets, dog, kittens                          & Pets                      \\
      3,402                        & flags, flag, redflags, red, blaring                     & (Red/Green) Flags               \\
      \bottomrule    
  \end{tabularx}
  \caption{Top 10 Topics of Female Dating Strategy.}
\label{tab:fds_top10}
  \end{table}

\begin{table}[t]
  \small
  \centering
    \setlength{\tabcolsep}{3pt}
    \begin{tabularx}{\columnwidth}{rL{5.2cm}L{2.2cm}}
      \toprule
{\bf Size} & {\bf Top 5 Words}                               & {\bf Label}  \\ 
      \midrule
        4,174                        & trufemcels, braincels, femcels, foreveralone, incels              & Femcels/Incels                  \\
        3,711                        & darker, ombre, brown, balayage, color                             & Hair Color          \\
        2,685                        & attractiveness, facially, conventionally, unattractive, lookswise & Attractiveness           \\
        2,229                        & europeans, ethnically, indians, southeast, arab                   & Ethnicity (Asia and Europe)         \\
        1,899                        & ugly, looksmaxxed, lookism, unattractive, duckling                & Lookism          \\
        1,722                        & ethnically, europeans, southeast, asians, asian                   & Ethnicity   \\
        1,516                        & stacylite, becky, stacy, normie, beckies                          & Stacies-Beckies                             \\
        1,509                        & ugliness, agoraphobia, friendless, looksmaxxed, agoraphobic       & Loneliness    \\
        1,464                        & moids, moid, jfl, foids, scrotes                                  & Moids \\
        1,437                        & internships, semester, undergrad, gpa, internship                 & College                          \\
        
        \bottomrule    
    \end{tabularx}
    \caption{Top 10 Topics of Femcels.}
  \label{tab:femcels_top10}
    \end{table}  
 
\descr{Mainstream Feminism.}
We observe that the posts in the sex conviction topic are generally oriented around discussions related to false accusations, e.g:
\begin{quote}
  \textit{Fake accusations are increasingly being used as a means of political manipulation.}
\end{quote}

Posts in the career, kyriarchy (a term linked to patriarchy and refers to an individuals's status in different social categories~\cite{abrahams2005critical,schussler2007power}), and toxic masculinity topics are generally about women's disadvantages due to patriarchal and toxic-masculine mindsets, e.g.:
\begin{quote}
  \textit{Women get more college degrees but men dominate high-paying jobs.}
\end{quote}

Posts in the elections topic are US presidential election oriented, focusing on Democratic candidates (e.g., Hillary Clinton and Bernie Sanders), %
while those in the sexual assault topic are generally about date rapes or discussions around intoxication and consent, e.g.:
\begin{quote}
  \textit{Sexual activity with a person unable to consent due to intoxication is considered rape.}
\end{quote}

We find two topics related to the Men's Right Activists (MRA) are among the top 10 topics of Mainstream Feminism.
Posts in the MRA (Reddit) topic are generally about the Mainstream Feminist subreddits' tensions between MRAs, e.g.:
\begin{quote}
  \textit{It seems that r/mensrights has come here for a downvote brigade.}
\end{quote}

\noindent We also see discussions on MRA and feminism relationship in the MRA topic, with participation from both sides, e.g.:

\begin{quote}
  \textit{I hold the belief that if MRA and Feminists were to collaborate, it could be a highly effective path towards achieving equality.}
\end{quote}

Overall, while certain Mainstream Feminist users oppose MRA on account of the toxic subgroups associated with the movement, there are others who advocate for strengthening the relationship between feminism and MRA as a means of advancing gender equality.

Aligning with feminist history~\cite{markowitz1990abortion}, the posts in the abortion topic are supportive of abortion rights.
In addition, the topic of fetuses in relation to abortion often involves ethical, moral, and legal debates, and the prevailing sentiment in these discussions tends to support the idea of abortion being permissible until a certain point in the pregnancy, e.g.:
\begin{quote}
  \textit{Once a fetus is considered a person, its right to live takes precedence over a woman's right to make choices regarding her own body, but prior to that, I have no objections.}
\end{quote}

\descr{Gender-Critical Feminism.}
Similar to Mainstream Feminism, GCF mostly talks about US presidential election in the elections topic, with focusing on Democratic candidates.

In the bathroom access topic, GCF mostly talks about sharing the bathroom with transgender people and men.
We see many posts criticizing the idea of separating restrooms as ``men's'' and ``others.''
While there are some posts that approach the issue of bathroom access for transgender individuals with a sense of fairness and common sense, there are also posts that display transphobia on this matter, e.g.:

\begin{quote}
  \textit{Opposite sex in women's bathroom/locker is not acceptable. Regardless of how much they ``pass''.}

\end{quote}

For trans-athletes, we see posts about the advantages trans-athletes might have in women's sports, with more leaning towards trans-exclusion compared to bathroom access, e.g.:

\begin{quote}
  \textit{The solution is not to allow them to participate in sports, rather they can seek employment in a different field.}

\end{quote}

In the clothing topic, we see many posts on gender-oriented discussions.
Even though we see posts with positive viewpoints on wearing clothes associated with the opposite gender, we also encounter transphobic ones, e.g.:
\begin{quote}
  \textit{No one is really transgender. A man in a dress and wig is still not a woman. Men are not women.}
\end{quote}

GCF mostly talks about bisexuality in the sexual orientation topic, where many posts have users identify themselves as bisexual.
In the leftism and sex work topics, we see support for leftism and opposition to sex work, as it aligns with the radical feminist perspective~\cite{willis1984radical,bromberg1997feminist}.

We also find discussions related to autogynephilia, a controversial term used for defining male's tendency to sexually arouse by thinking of themselves as woman~\cite{blancard1991clinical}.
We see many posts related to intersex individuals in the karyotype (an individual’s complete set of chromosomes~\cite{karyotype}) topic.
However, we again encounter many posts that approach the karyotypes in a transphobic manner, e.g.:
\begin{quote}
  \textit{Only those who were born with ambiguous genitalia and were misdiagnosed by physicians should be valued as trans women.}
\end{quote}

As yet another topic of discussion related to the potential challenges faced by transgender individuals, we find discussions related to gender dysphoria (a condition where individuals experience significant distress or discomfort due to a mismatch between their gender identity and the sex they were assigned at birth, or the physical characteristics associated with that sex~\cite{clinic2022}) in the respective topic.

\descr{Female Dating Strategy.}
True to their name, FDS's most popular topics are primarily oriented around dating and relationships.
In the dating apps, valentines day, jewelry, orgasm, blocking/unblocking, restaurants, and (red/green) flags topics, FDS share their romantic and sexual experiences.
Many of the posts in these topics have leaning to categorize the partners based on their ability to provide, e.g.:  
\begin{quote}
  \textit{Gift giving in a romantic relationship signals feelings. A thoughtful gift is a green flag, a mediocre effort is yellow, and low effort is a red flag. }
\end{quote}
We see lots of acronym use, e.g., hvm for high-value males and lvx for low-value ex (partner) in posts, which can be a sign of a toxic culture of FDS.
In the housework topic, FDS mostly talk about not doing housework for their partners or letting their partners do the housework, e.g.:
\begin{quote}
  \textit{He is a LVM. He has driven his girlfriends away by expecting them to clean up after him, and his living space is always dirty.}
\end{quote}

As another indicator of FDS's toxicity, several posts stereotype ethnicities in the ethnicity topic, e.g.:
\begin{quote}
  \textit{It is harder for white men to be HV, compared to other races.}
\end{quote}

In the pets topic, many posts evaluate the partners of FDS users based on their approaches to pets/animals, e.g.:
\begin{quote}
  \textit{The way he treats cats worse compared to dogs indicates that he’s a misogynist.}
\end{quote}

\begin{table}[t]
  \small
  \centering
    \setlength{\tabcolsep}{3pt}
    \begin{tabularx}{\columnwidth}{rL{5.5cm}L{1.6cm}}
      \toprule
{\bf Size} & {\bf Top 5 Words}                               & {\bf Label}  \\ 
      \midrule
      8,655                        & trp, theredpill, rpw, asktrp, rp                            & The Red Pill               \\
      6,714                        & nagging, emotions, stonewalling, nag, accusatory            & Emotions                        \\
      5,320                        & girlwithabike, rpws, trpw, mssadiedunham, introspective     & Red Pill Women                       \\
      4,942                        & captain, deferring, captains, defer, surrendering           & Submissiveness                         \\
      4,450                        & exclusivity, commitment, ltrs, ltr, plates                  & Relationships                    \\
      3,571                        & cohabitation, cohabitating, engaged, timelines, engagement  & Engagement                           \\
      3,549                        & chicken, sauce, cheese, rice, potatoes                      & Cooking                   \\
      3,305                        & leggings, jeans, blouse, blouses, cardigan                  & Clothing                     \\
      3,171                        & alphas, betas, alpha, beta, sigma                           & Alpha-Beta Masculinity                     \\
      3,091                        & mba, undergrad, internships, bachelors, majors              & Education                       \\      
      \bottomrule    
  \end{tabularx}
  \caption{Top 10 Topics of Red Pill Women.}
\label{tab:rpw_top10}
  \end{table}  

\begin{table}[t]
  \small
  \centering
    \setlength{\tabcolsep}{3pt}
    \begin{tabularx}{\columnwidth}{rL{5.5cm}L{2cm}}
      \toprule
{\bf Size} & {\bf Top 5 Words}                               & {\bf Label}  \\ 
      \midrule
        1,588                       & independence, financially, fulfillment, contentment, gtow         &  Financial Independence             \\
        1,052                       & wgtow, mgtow, gtow, mgtows, blackpill                             &  WGTOW/ MGTOW               \\
        391                        & friendships, platonic, friendship, companionship, bonds           &  Friendship               \\
        362                        & rental, renting, utilities, rent, condo                           &  Renting               \\
        292                        & evolutionarily, offspring, reproduce, species, procreate          &  Evolution              \\
        290                        & retirement, savings, investments, mortgage, retire                &  Financing              \\
        273                        & brigading, modmail, admins, brigaded, mod                         &  Moderation               \\
        258                        & sons, mothers, grandchildren, fathers, daughters                  &  Family              \\
        238                        & adoption, childfree, surrogate, adoptive, procreating             &  Adoption               \\
        233                        & garden, plants, gardening, trees, planting                        &  Plants              \\
       
        \bottomrule    
    \end{tabularx}
    \caption{Top 10 Topics of Women Going Their Own Way.}
  \label{tab:wgtow_top10}
    \end{table}

\descr{Femcels.}
We find that the posts in  femcels/incels topic are centered around being a femcel, and opposed to incels, e.g.:
\begin{quote}
  \textit{Try to imagine valuing the opinions of incels.}
\end{quote}
In general, Femcels talk about their appearance more negatively than the other communities.
In the hair color and college topics, Femcels mostly share their experiences and give/seek advice.
Additionally, we observe that in the college topic, Femcels share their negative experiences, often attributing them to their appearance, e.g.:
\begin{quote}
  \textit{Even my sorority sisters wouldn't let me in on their group photos because of my look.}
\end{quote}
This can be related to the bullying and social rejection femcels typically face during their school and work lives~\cite {serrano2022}.

We also observe the impact of negative experiences faced by femcel users due to their appearance in various other topics, with many posts in the attractiveness, lookism, and loneliness topics in this direction, e.g.:
\begin{quote}
  \textit{I'm currently content with being a femcel, and I plan to use this time to work on improving my appearance to better align with societal norms.  }
\end{quote}

Femcels use Stacy to define attractive women and Becky as women who can charm a partner even though she is not as attractive as Stacies~\cite{colombo2022}.
Many posts in Stacies-Beckies are about defining ``them'', e.g.:
\begin{quote}
  \textit{Stacies are very attractive (8-10 over 10), Beckies are 5-7, and the rest might be marked as femcels.}
\end{quote}

Similar to FDS, we find many posts stereotyping ethnicities instead of defining individuals, e.g.:
\begin{quote}
  \textit{Honestly, you're worse than blacks.}
\end{quote}

Femcels converted an incel term, foid (deragotory term to indicate women are sub-human~\cite{jones2020incels}), to ``moid'' to use it as the same meaning for men~\cite{lucy2022discovering}.
Posts in this topic are usually about complaints from ``moids.''

\descr{Red Pill Women.}
We see discussions regarding TRP and being a RPW in The Red Pill and Red Pill Women topics.
In general, RPW users admire TRP, although many posts emphasize the differences between TRP and RPW, e.g.:
\begin{quote}
  \textit{We follow TRP ideology and use its theories and concepts to help achieve the goals related to women that are part of the red pill movement.}
\end{quote}

We see RPW sharing their experiences and giving each other advice in the emotions and relationships topics, with posts mentioning the term ``plate,'' a term used by TRP to define ``using women only for sex''~\cite{dignam2019misogynistic}, e.g.:
\begin{quote}
  \textit{Don't settle for being a plate to get commitment in hopes of obtaining commitment from someone, but it's not impossible for that commitment to happen either. }
 \end{quote}
In the submissiveness topic, RPW discuss about being submissive, where they refer to their partners as ``captains,'' e.g.:
\begin{quote}
  \textit{I'm owned. My Captain has the authority to establish my position in the way that he decides, and I have given my consent to all of them with pleasure.}
\end{quote}

In contrast to GCF, RPW tends to focus primarily on femininity in their discussions regarding clothing.
Although the cooking topic is primarily focused on sharing cooking recipes, a considerable number of posts emphasize that the authors are preparing meals for their husbands, partners, or ``captains.''
We see posts in the engagement topic are discussions related to engagements/marriages.
We also find RPW classify males into alphas and betas, similar to the Manosphere~\cite{ging2019alphas}, while RPW talk about their education and give/seek advice in the respective topic, where we also see many posts related to the education degrees of the husbands/partners of RPW.

\descr{WGTOW.}
Posts in the topics of WGTOW are mostly related to living a single lifestyle.
In the financial independence, renting, and financing topics, posts are mostly about being a WGTOW, emphasizing their struggles, e.g.:
\begin{quote}
  \textit{She advises me to be financially independent before considering marriage. She tried to be a good girl, but it did not work for her.}
\end{quote}

In the MGTOW/WGTOW topic, discussion is mainly about opposing MGTOW, and about being a WGTOW, e.g.:
\begin{quote}
  \textit{MGTOW have negative feelings towards women, while WGTOW have positive feelings towards women.}
\end{quote}

In the moderation topic, posts are generally about subreddit/user bans and discussions about subreddit mods, with posts about adopting a child as a part of the WGTOW lifestyle in the adoption topic.
In the evolution topic, discussions include justifying sexual instincts for the WGTOW lifestyle, e.g.:
\begin{quote}
  \textit{It's simply instinct to reproduce. Nobody likes to recognize that animal instincts have influence over them, so they make up ludicrous explanations to justify their basic desires.}
\end{quote}

Finally, WGTOW users seem to share their experiences and thoughts with friendship, family, and plants topics.

\begin{figure}[t]
	\centering
	\includegraphics[width=\columnwidth]{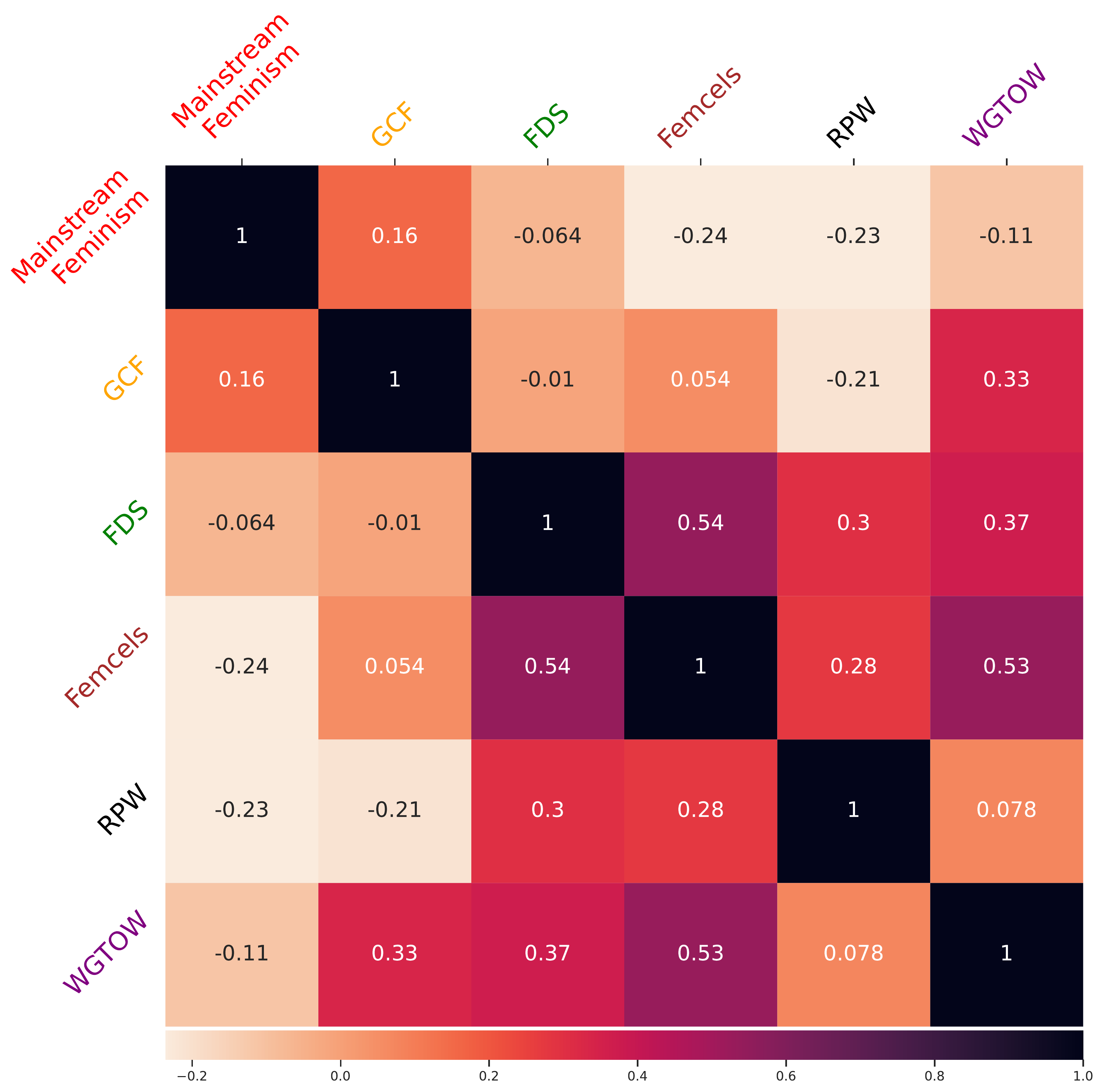}
	\caption{Cosine distances between Doc2Vec vector centroids of each community. Higher scores (darker color) mean posts in two communities are more similar.
	}
	\label{fig:document_similarity}
\end{figure}

\descr{Document Similarities.}
We compare the general leaning of posts of each community by looking at the pairwise cosine similarities between the centroids of the Doc2Vec vectors of their posts (See Figure~\ref{fig:document_similarity}).
GCF posts are most similar to Mainstream Feminism, while WGTOW ones are more than twice as similar to GCF compared to Mainstream Feminism.

We also find Femcels are most similar to FDS (cosine similarity: 0.54) and WGTOW (cosine similarity: 0.53) communities, whereas, FDS posts are most similar to Femcels. 
Similar to Figure~\ref{fig:normalized_user_overlaps}, we find low similarity between RPW and other Manosphere Analog communities.

\descr{Takeaways.}
Mainstream Feminism is arguably more focused on problems related to the patriarchy and toxic masculinity, while GCF also concentrates on issues related to transgender people, with some of them clearly crossing the threshold into transphobia.
Aligning with their user-base similarities (see Section~\ref{sec:user_analysis}), we find FDS, Femcels, and WGTOW are more similar to each other than RPW.

Additionally, while Femcels and WGTOW express opposition to their respective Manosphere counterparts, RPW expresses support for their analogs. 
This is also evident in their posts, where RPW users express more admiration for the opposite sex, often due to traditional family roles, compared to the other Manosphere Analogs.

\section{Toxicity Analysis}
\label{sec:perspective}

Since six subreddits from GCF and Femcels were banned due to hateful content, we set out to compare the differences in toxicity levels across the communities.
Ribeiro et al.~\cite{horta2021platform} show that a Manosphere community, r/Incels, does not significantly increase their toxicity level after migrating to incels.co.
We also measure the differences in toxicity levels of GCF and Femcels before and after their migrations to their own platforms, namely, ovarit.com and thepinkpill.co.
These platforms have a similar structure to Reddit and refer to their ``subreddits'' as channels.

\descr{ovarit.com.}
We collect 658,962 posts between July 2020 and February 2022 using a custom crawler.
Table~\ref{tab:ovarit_top} reports the top 20 most popular channels of ovarit.com.
GenderCritical is the most popular channel with 287,780 posts, which accounts for 43.6\% of all posts of ovarit.com.
We also see many Radical Feminist/Gender-Critical Feminist (e.g., ItsAFetish) Reddit communities on this platform.

\descr{thepinkpill.co.}
We collect submissions (and their comments) between February 2021, the month this platform was founded, and February 2022.
Overall, we gather 26,778 posts (submissions + comments) from 128 channels.
As reported in Table~\ref{tab:thepinkpill_top}, TruFemcels is the most popular channel with 8,380 posts, which makes up 31.2\% of all posts.
We also find other femcel subreddits (Vindicta, AskTrueFemcels, and PinkpillFeminism), and FemaleDatingStrategy among the top 20 channels.
WGTOW and TheGlowUp are also on this platform, as the 23rd and 30th most popular channels, respectively.
This makes thepinkpill.co is not just a hub for Femcels, but also for the other Manosophere Analogs, except for RPW.

\begin{table}[t]
  \small
  \centering
  \begin{tabular}{rlr}
      \toprule
      {\bf Rank} & {\bf Channel} &  {\bf  \#Posts}\\
      \midrule
      1                        & GenderCritical             & 287,780           \\
      2                        & Radfemmery                 & 72,387            \\
      3                        & TransLogic                 & 65,132            \\
      4                        & Women                      & 59,252            \\
      5                        & WomensLiberation           & 54,978            \\
      6                        & ItsAFetish                 & 9,767             \\
      7                        & NameTheProblem             & 8,771             \\
      8                        & Books                      & 7,158             \\
      9                        & Activism                   & 6,942             \\
      10                       & Lesbians                   & 6,696             \\
11                       & FeministVideos             & 6,618 \\
12                       & SaveWomensSports           & 6,104            \\
13                       & WomensHealthLounge         & 5,855           \\
14                       & Television                 & 5,762             \\
15                       & TerfIsASlur                & 5,689        \\
16                       & GoodNewsForWomen           & 5,259             \\
17                       & WomensHistory              & 4,844      \\
18                       & Cancelled                  & 4,516        \\
19                       & STEM                       & 3,862        \\
20                       & Movies                     & 3,859          \\

      \bottomrule    
  \end{tabular}
  \caption{Most popular channels of ovarit.com. }
\label{tab:ovarit_top}
  \end{table}

\begin{table}[t]
  \small
  \centering
  \begin{tabular}{rlr}
      \toprule
      {\bf Rank} & {\bf Channel} &  {\bf  \#Posts}\\
      \midrule
        1                        & TruFemcels             & 8,380                \\
        2                        & General                & 8,276                \\
        3                        & ThePinkPill            & 1,776                \\
        4                        & PinkPilledNormies      & 783                  \\
        5                        & FemaleDatingStrategy   & 610                  \\
        6                        & Vent                   & 399                  \\
        7                        & OffMyChest             & 344                  \\
        8                        & Vindicta               & 330                  \\
        9                        & TheMaleWall            & 323                  \\
        10                       & AskTruFemcels          & 289                  \\
11                       & PickMesNeverProsper    & 259 \\
12                       & Memes                  & 249       \\
13                       & Music                  & 238      \\
14                       & NameTheProblem         & 229      \\
15                       & PinkPillFeminism       & 222 \\
16                       & PenisPassDenied        & 221       \\
17                       & HardlineRadicalFeminism & 214 \\
18                       & MaleNature              & 198  \\
19                       & PinkPillDebate          & 176  \\
20                       & UnpopularOpinion        & 147      \\

        \bottomrule    
    \end{tabular}
    \caption{Most popular channels of thepinkpill.co.}
  \label{tab:thepinkpill_top}
    \end{table}  

\begin{figure*}[t]
	\centering
	\includegraphics[width=1.7\columnwidth]{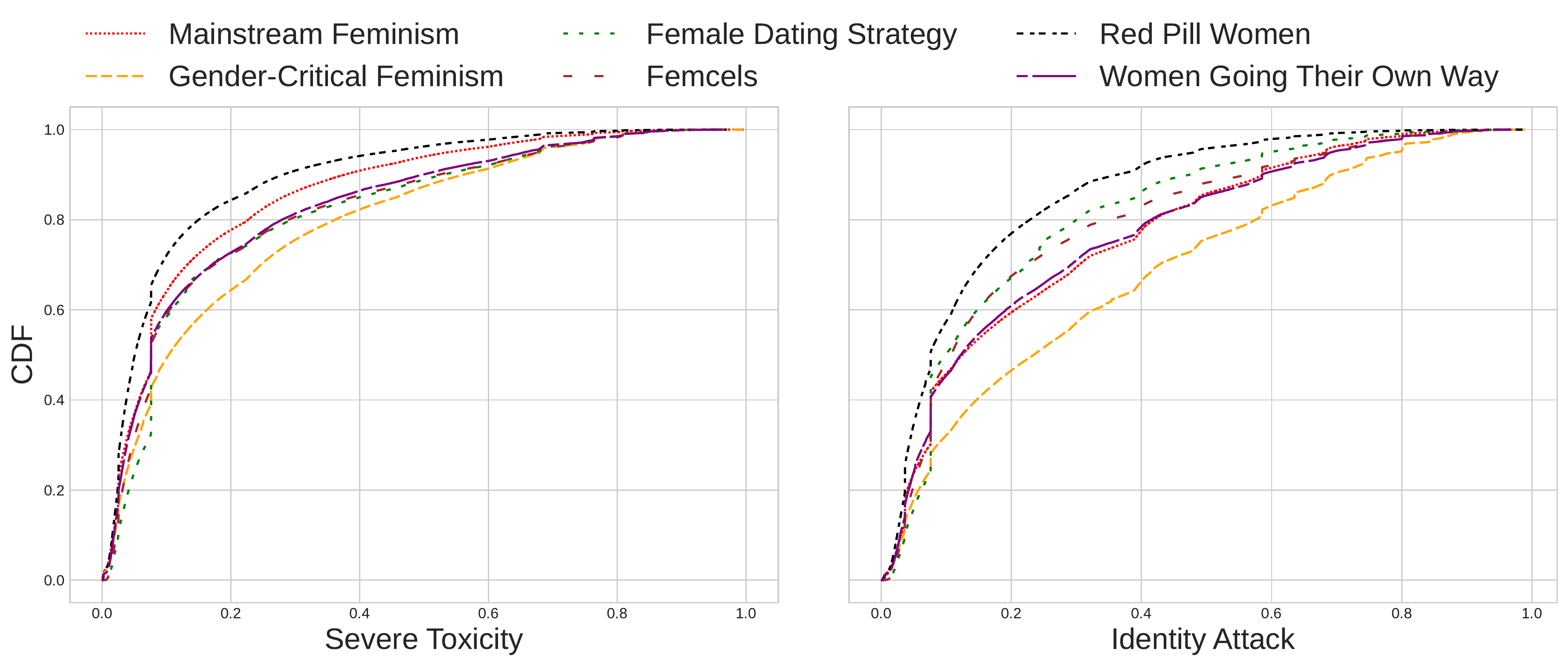}
	\caption{CDF of  severe toxicity scores and identity attack scores of each community (Reddit). Femcels, GCF, and WGTOW are the most toxic communities.}
	\label{fig:perspective_cdfs}
\end{figure*}

\begin{figure*}[t]
	\centering
	\includegraphics[width=1.7\columnwidth]{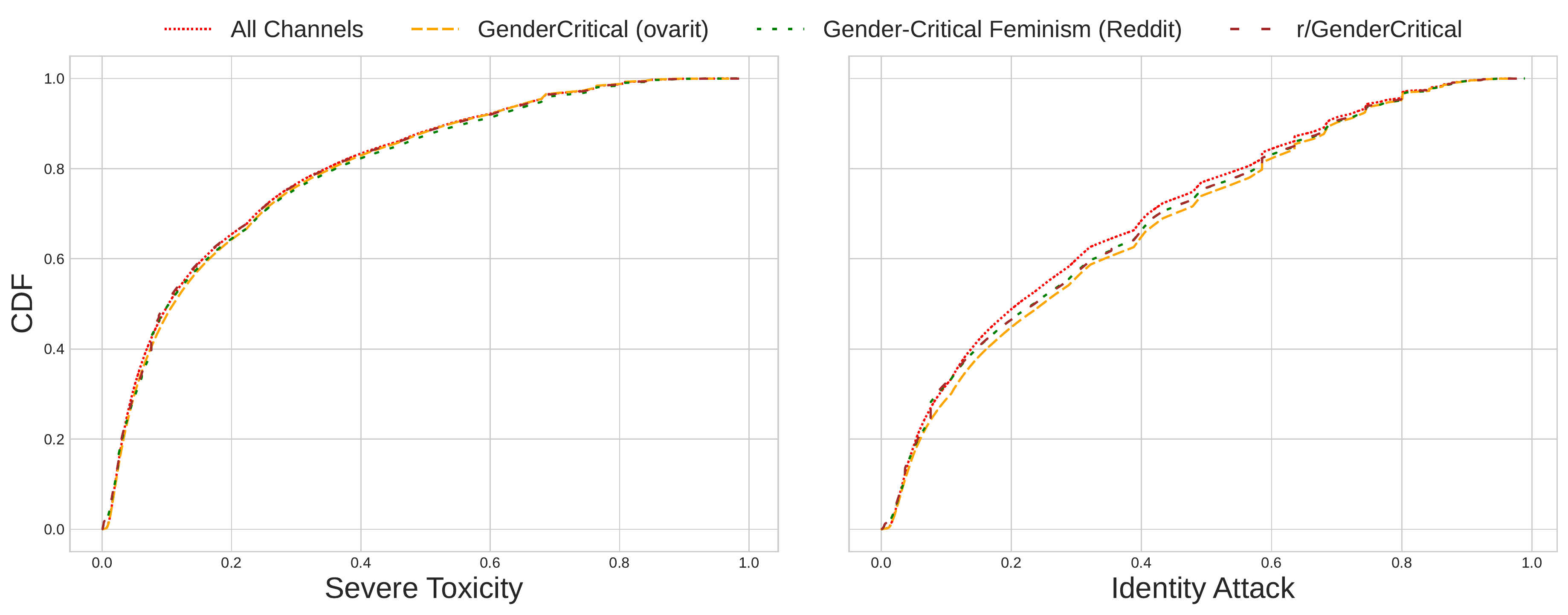}
  \caption{CDF of  severe toxicity scores and identity attack scores of ovarit's GenderCritical channel, Reddit's GCF community, r/GenderCritical, and the entire ovarit.com. GCF does not increase its toxicity levels after being deplatformed.}
  \label{fig:ovarit_perspective_cdfs}
\end{figure*}

\begin{figure*}[t]
	\centering
	\includegraphics[width=1.7\columnwidth]{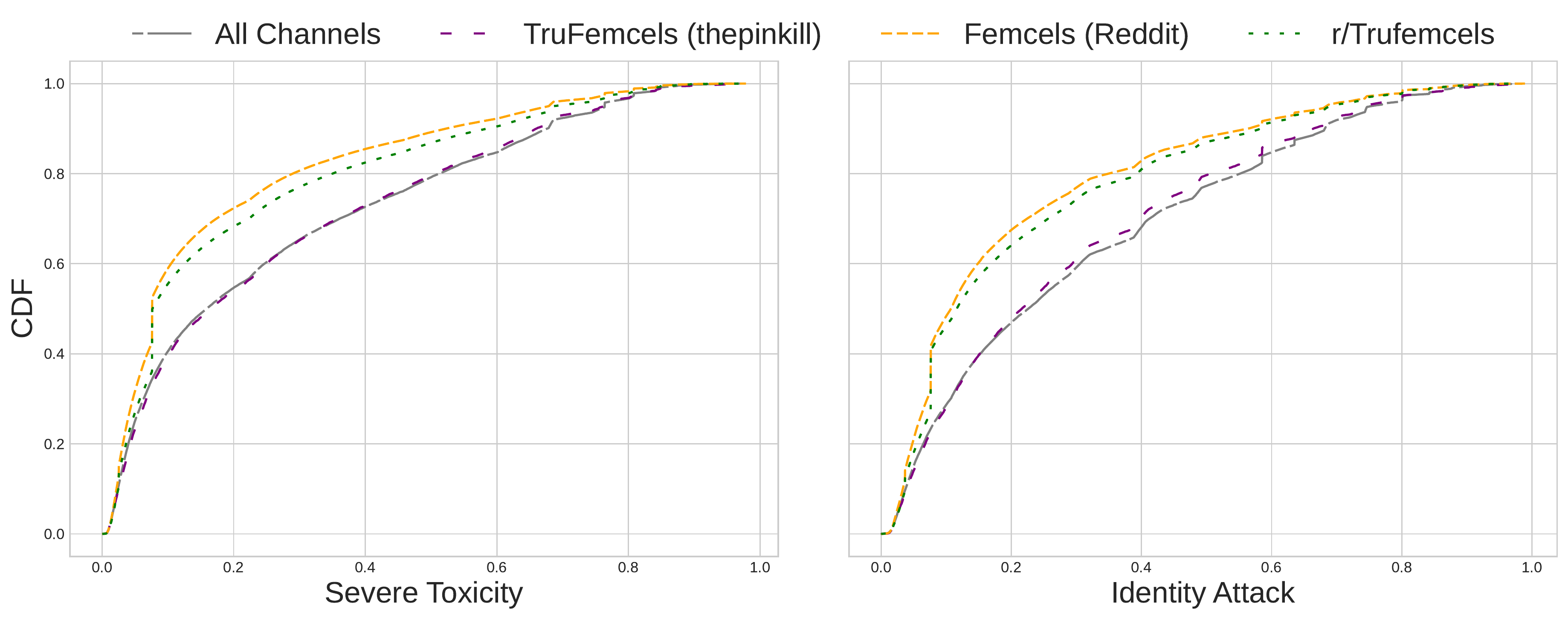}
  \caption{CDF of severe toxicity and identity attack scores of pinkpillco's TruFemcels channel, Reddit's Femcels community, r/Trufemcels, and the entire thepinkpill.co. Femcels increased their proportion of posts with high toxicity after their migration, alongside with their high identity attack proportions in their posts.}
  \label{fig:pinkpillco_perspective_cdfs}
\end{figure*}

\descr{Measuring Toxicity Levels.}
We use Google's Perspective API~\cite{perspective}, which relies on several machine learning models to measure abusive content prpoviding scores from 0 to 1, trained on millions of comments using multiple Internet sources~\cite{rieder2021fabrics}. 
Even though it is a widely used tool in the research community~\cite{mittos2020and,gruzd2020studying,shen2022xing}, Perspective is not without limitations, with drawbacks including biases and the tendency to mislabel conversational patterns it has not been encountered during its training~\cite{perspectivefaq}.

In our analysis, we opt for the \texttt{SEVERE\_TOXICITY} model; we also use the \texttt{IDENTITY\_ATTACK} model, which measures negative or hateful content targeting someone because of their identity (see~\cite{perspective}).   
Following Hoseini et al.~\cite{hoseini2021globalization}, we consider scores $\geq 0.8$ as a \emph{high} score.
Since Ribeiro et al. did not measure the \texttt{IDENTITY\_ATTACK} scores of r/Incels and the platform they migrated (incels.co), we also measure their \texttt{IDENTITY\_ATTACK} scores, finding a slight increase (0.06\% to 0.09\%) after deplatforming.
A 2-sample KS Test shows that the difference in distributions are statistically significant for each pair for each Perspective API model for each platform ($p < 0.05$ for all pairs, all $p$ are adjusted for multiple testing using Benjamini-Hochberg method~\cite{benjamini1995controlling}).

\begin{table*}[t]
  \centering
\small
    \setlength{\tabcolsep}{3.7pt}
    \begin{tabular}{lrr|lrr|lrr|lrr}
        \toprule
        \multicolumn{3}{c|}{\bf Gender-Critical Feminism}                                                          &  \multicolumn{3}{c|}{\bf ovarit.com}             & \multicolumn{3}{c|}{\bf Femcels}                       & \multicolumn{3}{c}{\bf thepinkpill.co}            \\
        \midrule
        \bf Named Entity                               & \bf \%High      &  \bf\#Posts               & \bf Named Entity    & \bf\%High & \bf\#Posts           & \bf Named Entity     & \bf\%High & \bf\#Posts                & \bf Named Entity     & \bf\%High & \bf\#Posts          \\    
        \midrule
        Transbians                                 & 61.58       & 151                   & Transbians     & 52.43    & 82              & Muslims         & 52.76  & 163                    & Islam           & 53.33   & 15                                                            \\
        Lesbians                                   & 51.36       & 220                   & Bill Burr      & 48.64    & 37              & Muslim          & 43.13  & 561                    & Muslim          & 52.63   & 19                                                        \\
        Grindr                                     & 42.55       & 282                   & Islamophobia   & 42.30    & 52              & Islam           & 43.10  & 342                    & LGBT            & 46.15   & 13                             \\
        Lesbian                                    & 41.45       & 441                   & Lesbians       & 42.10    & 153             & Bbc             & 32.07  & 106                    & Nazi            & 44.44   & 18                               \\
        Muslims                                    & 38.91       & 1,814                  & Jews           & 35.61    & 452             & Jewish          & 31.60  & 174                    & Nazis           & 42.85   & 14                                              \\
        Mexicans                                   & 38.67       & 106                   & Girldick       & 35.38    & 65              & Islamic         & 30.09  & 103                    & Asian           & 31.70   & 82                           \\
        Islamophobia                               & 37.60       & 234                   & Grindr         & 32.35    & 68              & White           & 30.00  & 120                    & Christianity    & 30.76   & 13                                      \\
        WLW                                        & 35.87       & 132                   & Lesbian        & 31.90    & 163             & LGBT            & 28.38  & 155                    & Christian       & 29.62   & 27                \\
        Jews                                       & 35.83       & 1,013                 & Muslims        & 31.62    & 370             & Nazis           & 28.30  & 159                    & African         & 27.77   & 18            \\
        Non-Muslim                                 & 35.37       & 147                   & White          & 31.61    & 155             & Indians         & 27.65  & 236                    & Western         & 27.77   & 18        \\
   
        \bottomrule
    \end{tabular}
    \caption{ Top 10 entities of Gender-Critical Feminism, ovarit.com, Femcels, and thepinkpill.co that appeared more than their specific thresholds with high identity attack scores with their high score frequencies and total number of posts they appeared.}
    \label{tab:gcf_femcels_ner_perspective}
\end{table*}

\descr{Reddit.}
Figure~\ref{fig:perspective_cdfs} plots the cumulative distributions of \texttt{SEVERE\_TOXICITY} and \texttt{IDENTITY\_ATTACK} scores of each community.
For both models, GCF tends to have higher scores than the other communities.
We find 12.6\% of the posts of GCF have $\geq 0.5$  \texttt{SEVERE\_TOXICITY} score, followed by FDS ($11\% \geq 0.5$) and Femcels ($10.7\% \geq 0.5$).
However, we find Femcels have the highest percentage of high \texttt{SEVERE\_TOXICITY} scores with 1.71\% rate (GCF is second with 1.55\% and WGTOW is third with 1.51\%).
For \texttt{IDENTITY\_ATTACK}, GCF has higher scores for both posts with $\geq 0.5$ (24.2\%) and $\geq 0.8$ (4.63\%).
GCF is followed by WGTOW and Mainstream Feminism for scores $\geq 0.5$ (14.6\% and 14.1\%), and Femcels and WGTOW for scores $\geq 0.8$ (2.05\% and 2.04\%).
We also find RPW has the least tendency to post with high \texttt{SEVERE\_TOXICITY} and \texttt{IDENTITY\_ATTACK} scores compared to the other communities.

\descr{ovarit.com.}
We compare the \texttt{SEVERE\_TOXICITY} and \texttt{IDENTITY\_ATTACK} score distributions of the GenderCritical channel, the entire ovarit.com, subreddits of GCF, and r/GenderCritical -- see Figure~\ref{fig:ovarit_perspective_cdfs}.
We find Reddit's GCF has a higher proportion of \texttt{SEVERE\_TOXICITY} scores for both $\geq 0.5$ (12.6\%) and $\geq 0.8$ (1.55\%).
Although r/GenderCritical, the entire ovarit.com and its GenderCritical channel have relatively similar proportions for scores $\geq 0.5$ (11.82\%, 11.63\%, and 11.83\%) and $\geq 0.8$ (1.32\%, 1.28\% and 1.22\%).
However, for \texttt{IDENTITY\_ATTACK} scores, ovarit.com and its GenderCritical have slightly more proportion of posts for scores $\geq 0.5$ (25.65\% and 25.60\%) than Reddit's GCF and r/GenderCritical (24.25\% and 24.21\%).
Even though ovarit.com's GenderCritical has the highest proportion of high \texttt{IDENTITY\_ATTACK} scores (4.64\%), we find Reddit's GCF has a similar proportion (4.63\%), followed by r/GenderCritical and the entire ovarit.com (4.45\% and 4.14).

\descr{thepinkpill.co.}
We find \texttt{SEVERE\_TOXICITY} scores of thepinkpill.co and its TruFemcels channel are higher than Reddit's Femcels community and r/Trufemcels -- see Figure~\ref{fig:pinkpillco_perspective_cdfs}.
While more than 20\% of the thepinkpill.co and TruFemcels have scores $\geq 0.5$ (20.8\% and 20.3\%), this is 13.1\% for r/Trufemcels, and 10.7\% for Reddit's Femcels.
For high \texttt{SEVERE\_TOXICITY}, thepinkpill.co and TruFemcels have a 3.3\% and 3.2\% ratio, while we find 2.1\% for r/Trufemcels and 1.7\% for Reddit's Femcels.
Our findings show similar results to~\cite{horta2021platform}, where another gender-based ideological community, Femcels, does not decrease its toxicity levels after their migration due to deplatforming. 
We find thepinkpill.co has the highest proportion of \texttt{IDENTITY\_ATTACK} scores $\geq 0.5$ with 22.7\%, followed by TruFemcels (20.3\%), r/Trufemcels (12.9\%), and Reddit's Femcels (11.7\%).
The exact order preserves for high \texttt{IDENTITY\_ATTACK} scores with 3.7\%, 3.4\%, 2.2\%, 2.0\% respectively.

\descr{Which identities do GCF and Femcels attack?}
Given that the GCF and Femcels subreddits were banned from Reddit for engaging in hate speech and have the highest proportions of high \texttt{IDENTITY\_ATTACK} scores, we conduct an analysis to examine the entities targeted by these communities.
Additionally, we explore the entities targeted in ovarit.com and thepinkpill.co to gain insight into differences in their targeting behaviors following their deplatforming.

\descr{Named Entity Recognition.}
Next, we extract named entities using the \textit{en\_core\_web\_lg} model of the SpaCy library~\cite{spacy}, which has been trained on datasets like WordNet~\cite{miller1995wordnet} and Common Crawl~\cite{commoncrawl}.
\textit{en\_core\_web\_lg} is a widely used tool in the research community~\cite{papasavva2020raiders,algamdi2022twitter,wang2021multi}.

To gain a clearer understanding of the identities targeted by these communities, we narrow our focus to (named) entities that appear in posts more than 100 times on GCF and Femcels.
For ovarit.com and thepinkpill.co, we adjust these thresholds to account for the lower number of posts on the platforms to which these communities migrated.
Specifically, we scale the thresholds by the fraction of total posts on the migration platforms compared to the total number of posts on their respective origin communities, with keeping the minimum possible threshold as 10.

Table~\ref{tab:gcf_femcels_ner_perspective} reports the top 10 entities with high \texttt{IDENTITY\_ATTACK} scores for GCF, ovarit.com, Femcels, and thepinkpill.co.
Consistent with the findings in Section~\ref{sec:topic_analysis}, our analysis reveals that GCF predominantly targets transgender identities. 
While the entities ``lesbians,'' ``lesbian,'' and ``WLW (Women Love Women)'' may appear to suggest that GCF also targets lesbians, a closer examination of the posts containing these entities indicates that they are mainly associated with the conflict between trans-exclusionary lesbians and transgender individuals~\cite{worthen2022my}, e.g.:
\begin{quote}
  \textit{A lot of transgender-women harass lesbians since they do not find the male body sexually attractive.}
\end{quote}
We also encounter transphobic attitudes in GCF posts that mention  ``Grindr,'' a popular LGBTIQ+ dating app.
Besides transgender identities, GCF targets ``Muslims,'' ``Mexicans,'' and ``Jews,'' although the appearance of ``Islamophobia'' and ``non-Muslim'' may suggest that there is a polarization among GCF related to Islam.

We encounter a similar picture for ovarit.com, where the primary target of this platform is also transgender identities.
However, unlike GCF, ovarit.com also targets ``Bill Burr (comedian)'' and ``white'' entities more frequently than ``Mexicans'' and ``non-Muslim''.
Our manual inspection on posts that include Bill Burr suggests that GCF targets the comedian due to his statements regarding the role of white women in black oppression~\cite{kelley2020}, which also explains the appearance of ``white'' in our results -- e.g.:
\begin{quote}
  \textit{It's clear that Bill Burr doesn't give a shit about women of color and is simply trying to create conflict between women.}
\end{quote}

Our findings for Femcels are consistent with those presented in Section~\ref{sec:topic_analysis}, indicating that this community primarily directs its attacks towards ethnicities/identities, as evidenced by all the entities listed in Table~\ref{tab:gcf_femcels_ner_perspective} being associated with either ethnicities (BBC, Jewish, White, Indians) or identities (Muslims, Muslim, Islam, Islamic, LGBT).

Our analysis of thepinkpill.co reveals a similar pattern of targeting towards specific groups, such as ``Muslims,'' ``Asians,'' ``Christians,'' ``Africans,'' and ``westerners''.
Our examination of both Reddit and thepinkpill.co demonstrates that Muslims are the primary target for Femcels, e.g.:
\begin{quote}
  \textit{Ex-Muslims are the only good Muslims.}
\end{quote}

Overall, deplatforming does not seem to have altered the set of identities that GCF and Femcels primarily target.

\descr{Takeaways.}
Communities that are banned from Reddit because of their toxic behavior (Femcels and GCF) are indeed the most toxic communities of OWIS.
We find a decrease in the proportion of high toxicity scores for GCF after their deplatforming.
This is somewhat in contrast with previous work on the Manosphere showing that the toxicity levels of incels do not decrease after their migration~\cite{ribeiro2020evolution}. %
We also find an increase in Femcels, GCF, and incels in terms of high identity attack scores after migrating to their new platforms.
Finally, despite deplatforming, the groups of people mainly targeted by GCF and Femcels do not seem to have changed.

\section{Discussion \& Conclusion}

This paper present a data-driven analysis of online women's ideological spaces (OWIS).
We quantitatively derived a taxonomy of OWIS consisting of 14 subreddits, and analyzed over 6M comments along several axes.
Among other things, we found that Gender-Critical Feminism has user-base similarities with Femcels and Women Going Their Own Way, which might seems odd, but can be explained by Gender-Critical Feminism's roots in radical feminism, which is generally more ``opposed'' to men than other forms of feminism.

Gender-Critical Feminism and Femcels are the most toxic communities, which is in line with the fact that they were been banned from Reddit.
When exploring their migrations to custom-built platforms upon being banned, we observe an increase in toxicity for Femcels, but not for Gender-Critical Feminism. 
We also see an increase in identity attacks after migration for both GCF and Femcels, which is one of the specific behaviors that they were banned from Reddit for in the first place.
Similar to what previous work found with alt-right communities like r/The\_Donald and Gab~\cite{horta2021platform,ali2021understanding}, deplatforming is correlated with increases in negative behavior for GCF and Femcels.

\descr{Acknowledgements.}
This material is based upon work supported by the National Science Foundation under Grant No. IIS-2046590 and CNS-1942610.
Chen Ling was supported by the Meta Research PhD Fellowship Award.

\small
\bibliographystyle{abbrv}
\bibliography{bibliography}

\end{document}